\documentclass[runningheads]{llncs}
\usepackage[T1]{fontenc}
\usepackage{graphicx}
\usepackage{subfig}         
\usepackage{multirow}
\usepackage{booktabs} 
\usepackage{booktabs}
\usepackage[misc]{ifsym}
\usepackage{amsmath}
\usepackage{amssymb}

\newcommand{\corr}{(\Letter)}
\usepackage{mwe}
\newcommand*\samethanks[1][\value{footnote}]{\footnotemark[#1]}
\makeatletter
\def\@fnsymbol#1{\ensuremath{\ifcase#1\or\star\or\dagger\or
   {\star\star\star}\or \ddagger\or \mathchar "278\or
   \mathchar "27B\or \|\or **\or \dagger\dagger\or
   \ddagger\ddagger \else\@ctrerr\fi}}
\makeatother

\begin{document}

\title{BAHSD: Bridging the Long-tail Gap via Adaptive Distillation in Black-box Sequential Recommendation}

\titlerunning{Adaptive Distillation in Black-box Sequential Recommendation}


\author{Xi Zhou\inst{1,2}\thanks{These authors contribute equally to this work.} 
\and
Famin Wu\inst{3}\samethanks 
\and 
Mingming Li\inst{1} \corr 
\and 
Hui Wang\inst{1,2} 
\and 
Hongyue Zhang\inst{1,2} 
\and 
Jiao Dai\inst{1} 
\and 
Jizhong Han\inst{1} 
\and 
Tao Guo\inst{1}
}
\authorrunning{Xi Zhou et al.}

  
\institute{
Institute of Information Engineering, Chinese Academy of Sciences, Beijing, China 
\and
School of Cyber Security, University of Chinese Academy of Sciences, Beijing, China
\and
Beijing Institute for General Artificial Intelligence, Beijing, China \\
\email{\{zhouxi, limingming, wanghui4042, zhanghongyue, daijiao, hanjizhong, guotao\}@iie.ac.cn}
\email{sa614368@mail.ustc.edu.cn}
}
\maketitle       

\begin{abstract}

Sequential recommendation systems are widely adopted but often deployed as black-box APIs, which has driven recent interest in model extraction to replicate their capabilities locally. However, the long-tail distribution induces severe signal heterogeneity: dense head sequences trigger the solidification of teacher preference, biasing extraction toward local patterns, while sparse tail sequences yield flat, noisy predictions. Existing one-size-fits-all extraction overlooks this disparity, resulting in noise overfitting and suboptimal knowledge transfer.
We propose BAHSD, a black-box adaptive distillation framework that handles signal heterogeneity via a multi-scale consistency probing mechanism to implicitly quantify signal reliability. Based on this, an adaptive hierarchical objective is designed: dynamic-temperature KL divergence mitigates preference solidification for high-confidence signals, while ranking consistency and InfoNCE contrastive learning provide noise-robust enhancement for low-confidence signals.
BAHSD consistently outperforms baselines, achieving up to 4.98\% gain over the teacher and 80\%+ improvement on tail users, offering a plug-and-play solution for high-fidelity black-box recommendation extraction.

\end{abstract}

\section{Introduction}
Sequential recommendation systems (SRS) employ advanced Transformer models, such as SASRec \cite{kang2018self} and BERT4Rec \cite{sun2019bert4rec,li2020symmetric}, to achieve state-of-the-art performance. However, these models are typically served as black-box APIs, which restricts access to internal parameters and gradients. This limitation renders black-box model extraction, the process of transferring knowledge from a remote API to a local model, a critical technique for reducing query costs and enabling downstream customization.

\begin{figure}[tp]
\centering
\includegraphics[width=0.7\textwidth]{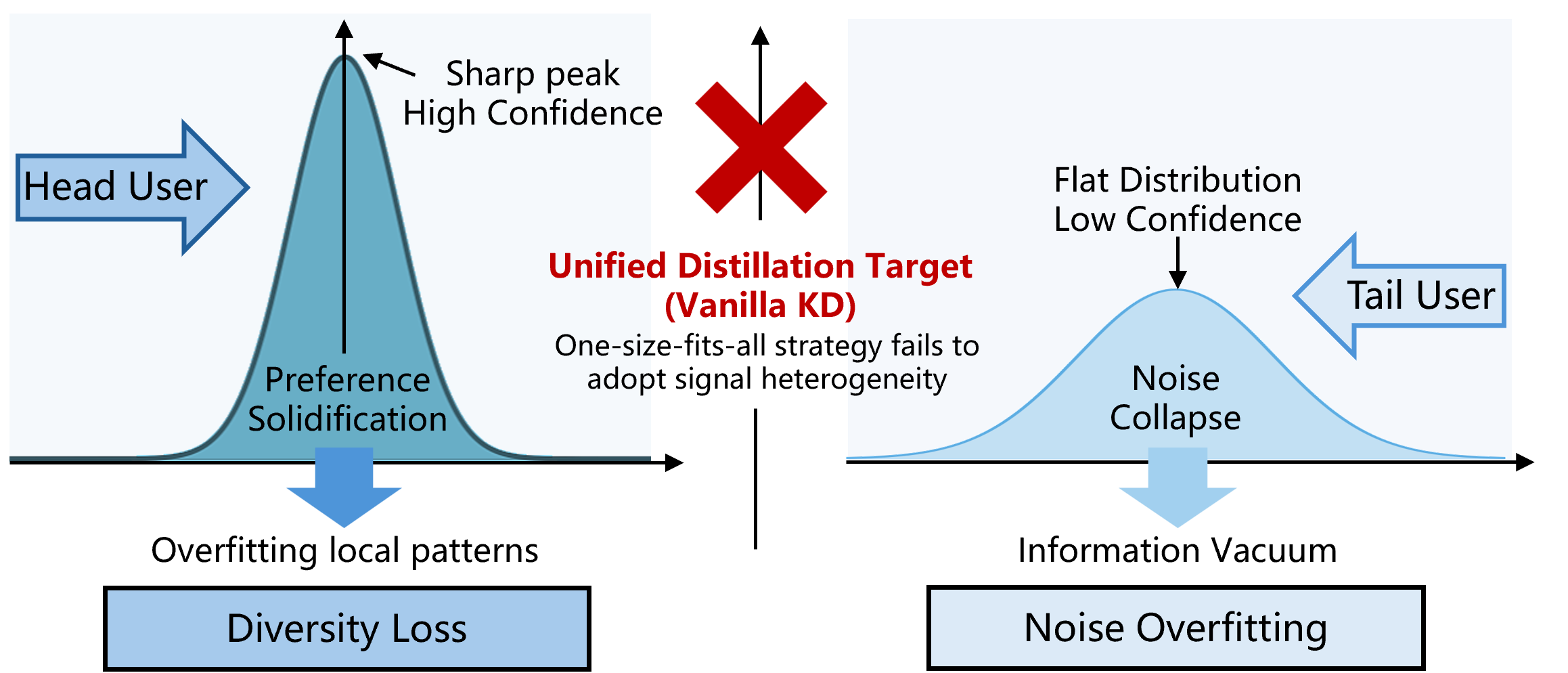}
\caption{Schematic diagram of supervision signal heterogeneity in teacher models.} 
\label{signal-hetero.png}
\end{figure}

Existing extraction methods, including UnKD \cite{chen2023unbiased} and RCE-KD \cite{zhu2025rejuvenating}, primarily focus on bias mitigation or distribution alignment under an implicit uniformity assumption that teacher signals are consistently reliable across all users. This assumption fundamentally conflicts with the long-tail distribution inherent in user interactions, thereby inducing severe \textbf{signal heterogeneity}. As illustrated in Fig.~\ref{signal-hetero.png}, teacher outputs exhibit two distinct degradation modes. First, \textbf{preference solidification} occurs in head users, where dense interactions yield sharp, low-entropy distributions that propagate biased preferences when mimicked by the student. Second, \textbf{information vacuum} arises in tail users, where sparse interactions produce flat, near-uniform distributions that degrade student performance through noise fitting. These modes are overlooked by one-size-fits-all distillation objectives, resulting in noise overfitting for tail users and underutilization of head signals. Moreover, traditional long-tail debiasing techniques rely on white-box access, such as gradients or embeddings, and are therefore inapplicable under black-box constraints. These observations motivate the development of an adaptive framework that can implicitly assess signal reliability and adjust distillation strategies accordingly.
To address this gap, we propose BAHSD (Black-box Adaptive Heterogeneous Signal Distillation), a framework that operates solely on teacher logits. BAHSD introduces a multi-scale consistency probing mechanism, which quantifies signal reliability by measuring the consistency of teacher outputs across subsequences of varying lengths (short, medium, and full) without requiring metadata or internal features. Based on the perceived reliability, an adaptive hierarchical objective is formulated. For high-confidence signals originating from head users, KL divergence with dynamic temperature scaling softens over-concentrated distributions to alleviate preference solidification. For low-confidence signals from tail users, ranking consistency combined with InfoNCE contrastive learning replaces aggressive distribution matching to enhance discriminative representations and suppress noise. A cross-scale symmetric KL consistency further regularizes the student to ensure robust predictions across different views, enabling end-to-end adaptive distillation without explicit user stratification or access to teacher internals.

Experiments on public benchmarks demonstrate that BAHSD consistently outperforms state-of-the-art baselines, achieving up to a 4.98\% improvement over the teacher model and boosting tail-user performance by over 80\% while maintaining head-user stability. BAHSD provides a model-agnostic, plug-and-play solution for high-fidelity black-box model extraction. Our contributions are threefolds:
\begin{itemize}
    \item We uncover and formalize signal heterogeneity in black-box sequential recommendation, identifying preference solidification for head users and information vacuum for tail users as two distinct degradation modes.
    \item We propose BAHSD, the first framework designed to adaptively handle this heterogeneity through multi-scale consistency probing and a reliability-aware hierarchical objective.
    \item Extensive experiments validate the effectiveness of BAHSD, demonstrating substantial gains for tail users and, in certain scenarios, performance that surpasses the teacher model under strict black-box constraints.
\end{itemize}
 
\section{Related Work}
\subsection{Sequential Recommendation}
Sequential recommendation captures users' dynamic preferences by modeling temporal dependencies in interaction histories \cite{boka2024survey,pan2026survey,li-etal-2024-shot,li2023adaptive,maconte,liqueryindex}. Transformer-based architectures, notably SASRec \cite{kang2018self} and BERT4Rec \cite{sun2019bert4rec}, have become the standard due to their ability to encode long-range patterns via self-attention. In practice, these models are often deployed as black-box APIs that expose only predictions while concealing internal parameters and gradients, necessitating distillation techniques that operate under such constraints \cite{gou2021knowledge,song2025knowledge}.

\subsection{Black-box Knowledge Distillation}
Knowledge distillation transfers knowledge from a teacher model to a student model \cite{hinton2015distilling}. Traditional distillation methods in recommendation systems follow white-box \cite{zhu2025exploring} or gray-box paradigms \cite{du2023ensemble,kang2020rrd,kweon2021bidirectional,liu2020general,zhu2025exploring}, which rely on intermediate features, attention maps, or gradient information that are inaccessible under black-box API settings. Recent black-box distillation efforts, including DHKD \cite{yang2025dual}, ABKD \cite{wang2025abkd}, ICD \cite{li2025logits}, and RLD \cite{sun2025knowledge}, primarily address loss conflicts and capacity gaps in general tasks such as image classification but lack recommendation-specific design. These methods treat teacher outputs as static distributions, overlooking both the temporal dynamics inherent in user sequences and the signal heterogeneity induced by long-tail user distributions. Approaches such as RCE-KD \cite{zhu2025rejuvenating}, Ekd4rec \cite{wang2025ekd4rec}, and DLLM2Rec \cite{cui2024distillation} attempt to refine ranking alignment through item partitioning, yet they operate at the output level without addressing the underlying quality disparity in teacher signals across different user groups.

\subsection{Long-tail Optimization in Recommendation}
Long-tail issues have been extensively studied in recommendation systems \cite{zhang2021model,han2024intra,wu2024coral,wang2025improving,zhang2021causal,wei2021model}. Methods such as HPSERec \cite{xu2025hpserec} and UnKD \cite{chen2023unbiased} focus on mitigating item-side popularity bias through stratified learning or unbiased distillation. Nevertheless, these methods exhibit two critical limitations: they primarily target item-side rather than user-side disparities, and they implicitly assume uniform teacher signal quality across all users. More fundamentally, their reliance on white-box access or explicit metadata renders them inapplicable under strict black-box constraints where only logits are available.

In contrast to prior work, we identify that the core bottleneck in black-box sequential recommendation lies in user-side signal heterogeneity—specifically, the dual degradation of teacher outputs into preference solidification for head users and information vacuum for tail users. BAHSD represents the first framework designed to address this problem under strict black-box constraints, achieving adaptive knowledge transfer through multi-scale consistency probing without requiring explicit user stratification or internal features.

\begin{table}[tp] 
\centering
\caption{Predictive reliability of teacher models across user tiers, with bold entries denoting optimal performance.}
\label{tab:performance_diff}
\small 
\renewcommand{\arraystretch}{1.2} 
\begin{tabular}{lccccc}
\toprule 
Dataset & Tier & \multicolumn{2}{c}{BERT4Rec} & \multicolumn{2}{c}{SASRec} \\
\cmidrule(lr){3-4} \cmidrule(lr){5-6} 
& & Recall@10 & NDCG@10 & Recall@10 & NDCG@10 \\
\midrule
\multirow{3}{*}{Amazon Beauty} & Head & \textbf{0.527} & \textbf{0.357} & \textbf{0.547} & \textbf{0.395} \\
& Mid  & 0.500 & 0.322 & 0.493 & 0.340 \\
& Tail & 0.478 & 0.308 & 0.474 & 0.322 \\
\midrule
\multirow{3}{*}{MovieLens-1M}  & Head & 0.689 & 0.478 & 0.713 & 0.487 \\
& Mid  & 0.745 & 0.549 & 0.791 & 0.562 \\
& Tail & \textbf{0.844} & \textbf{0.641} & \textbf{0.875} & \textbf{0.658} \\
\bottomrule
\end{tabular}
\end{table}

\section{Motivation Analysis}
\label{sec:motivation}
Existing black-box distillation methods for sequential recommendation assume uniformly reliable teacher signals, contradicting the intrinsic heterogeneity induced by long-tail user interactions. Our empirical analysis on Amazon Beauty \cite{ni2019justifying} and MovieLens-1M \cite{harper2015movielens} using SASRec and BERT4Rec reveals this discrepancy. We stratify users by sequence length into head (top 20\%), mid, and tail (bottom 60\%) groups, contrasting signal characteristics across tiers.
\subsection{Empirical Evidence of Signal Heterogeneity}
\begin{figure}[tp]
  \centering
  \subfloat[Amazon beauty on BERT4Rec]{
    \includegraphics[width=0.45\textwidth]{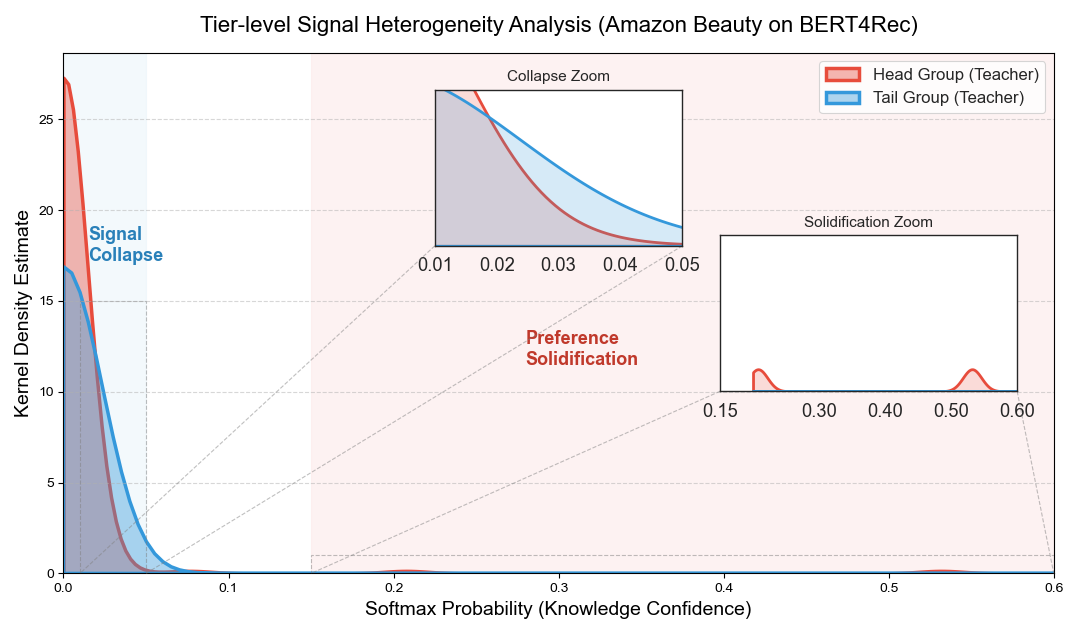}
    \label{subfig:beauty-bert}
  }
  \hfill  
  \subfloat[Amazon beauty on SASRec]{
    \includegraphics[width=0.45\textwidth]{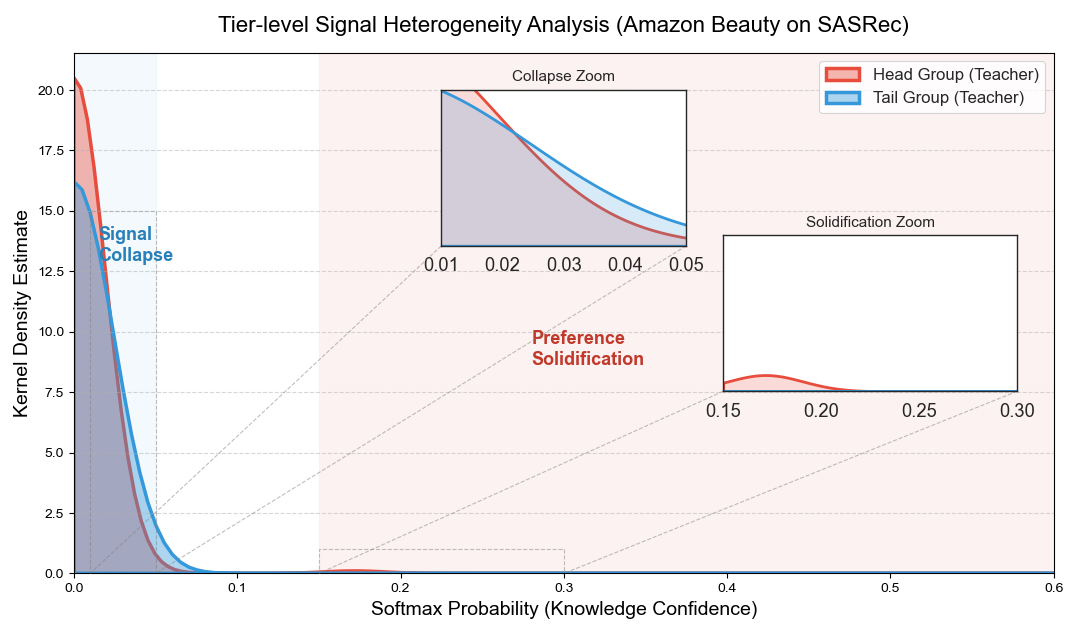}
    \label{subfig:beauty-sas}
  }\\[0pt]  
  \subfloat[MovieLens-1M on BERT4Rec]{
    \includegraphics[width=0.45\textwidth]{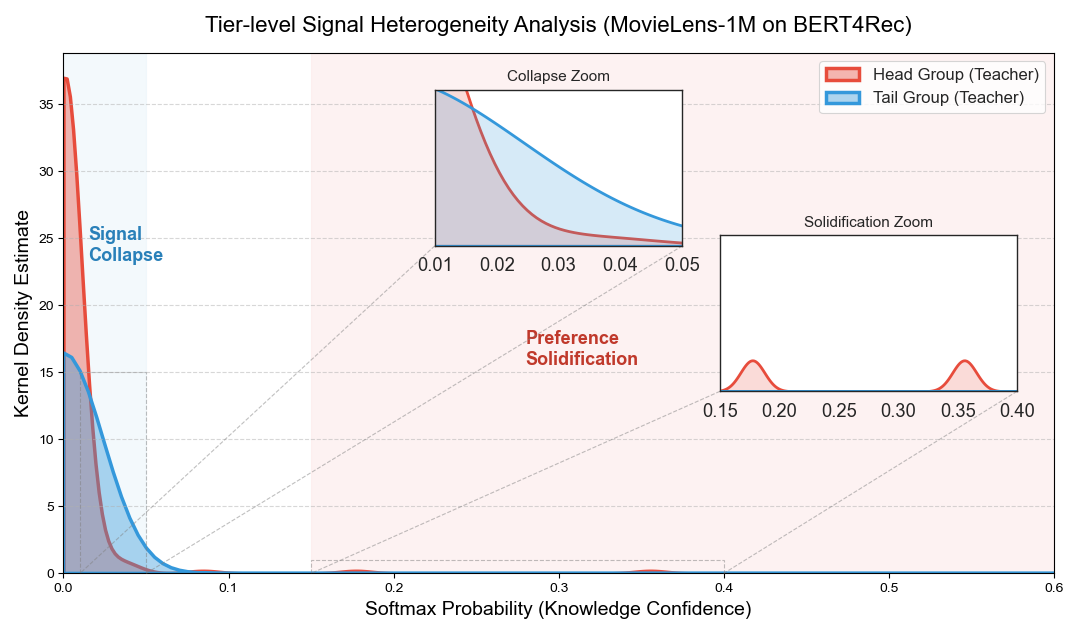}
    \label{subfig:ml1m-bert}
  }
  \hfill
  \subfloat[MovieLens-1M on SASRec]{
    \includegraphics[width=0.45\textwidth]{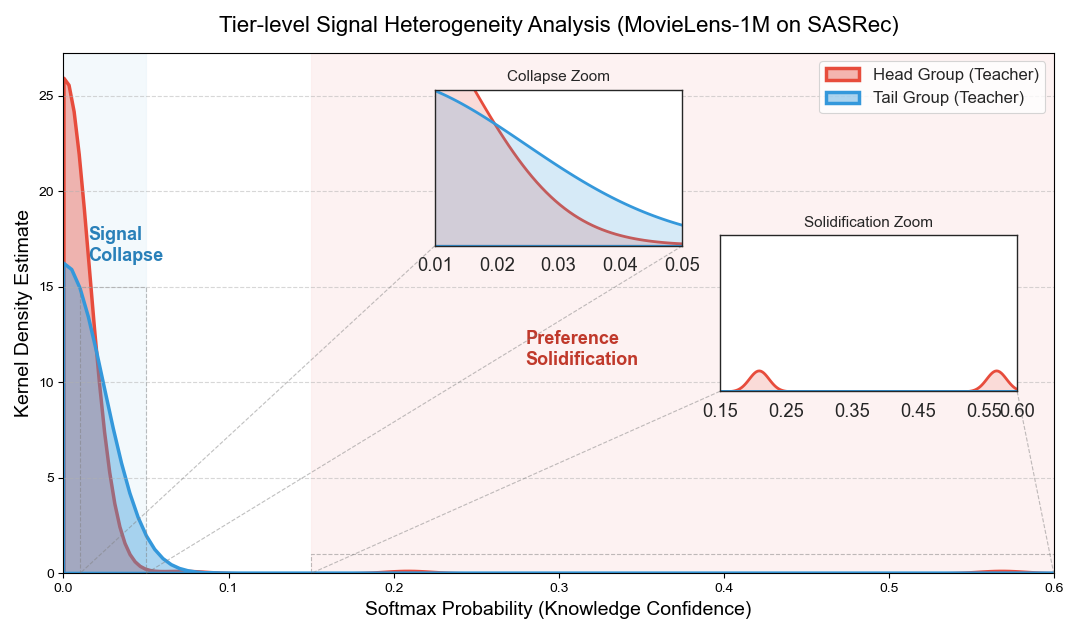}
    \label{subfig:ml1m-sas}
  }
  \caption{Logits distribution analysis on teacher models across user tiers.}
  \label{fig:reliable-anaysis}
\end{figure}

\textbf{Prediction performance analysis.} Table~\ref{tab:performance_diff} reports the performance of teachers at the user levels. On the dense MovieLens-1M dataset, we observe a striking performance inversion: tail users outperform head users, e.g., SASRec Recall@10: 0.875 vs. 0.713. This finding validates the phenomenon of \textit{preference solidification}, in which Transformer-based teachers overfit to long head sequences, thereby compromising generalization. For the sparse Amazon Beauty dataset, head-tier performance exceeds that of the tail as expected, although tail signals remain noisy and exhibit low confidence.
\textbf{Information-theoretic analysis.}  
Kernel Density Estimation (KDE) of the teacher logits shown in Fig.~\ref{fig:reliable-anaysis} reveals two distinct degradation modes. For tail users, a discriminative vacuum emerges where the probability mass concentrates near zero ($p<0.05$) with high entropy, indicating that signals are dominated by noise and rendering distribution matching ineffective. For head users, entropy-saturated solidification occurs where probability mass shifts non-linearly to high-confidence intervals ($p>0.15$). This low-entropy state reflects overconfidence and overfitting to local patterns.

\subsection{Theoretical Objective and Implicit Task Definition}

If an oracle could perfectly assess signal reliability $\mathcal{R}(u)$, the optimal stratified loss would take the form:
\begin{equation}
\min_{\theta_\mathcal{S}} \mathcal{L}_{Oracle} = \mathbb{E}_{u \in \mathcal{U}} \left[ \mathbb{I}(\mathcal{R}(u) \le \gamma_1) \mathcal{L}_{tail} + \dots + \mathbb{I}(\mathcal{R}(u) \ge \gamma_2) \mathcal{L}_{head} \right] 
\end{equation}
where $\gamma_1$ and $\gamma_2$ denote user stratification thresholds. However, explicit stratification is infeasible under black-box constraints. Therefore, our core motivation is to implicitly approximate this oracle by designing a distillation framework that perceives signal reliability and adaptively adjusts its optimization strategies without requiring metadata or internal access. BAHSD achieves this through multi-scale sequence probing and adaptive loss synergy, directly addressing the identified degradation modes.
\section{Methodology: The  Framework of BAHSD}
\label{sec:methodology}
\begin{figure}[tp]
    \centering
    \includegraphics[width=0.85\textwidth]{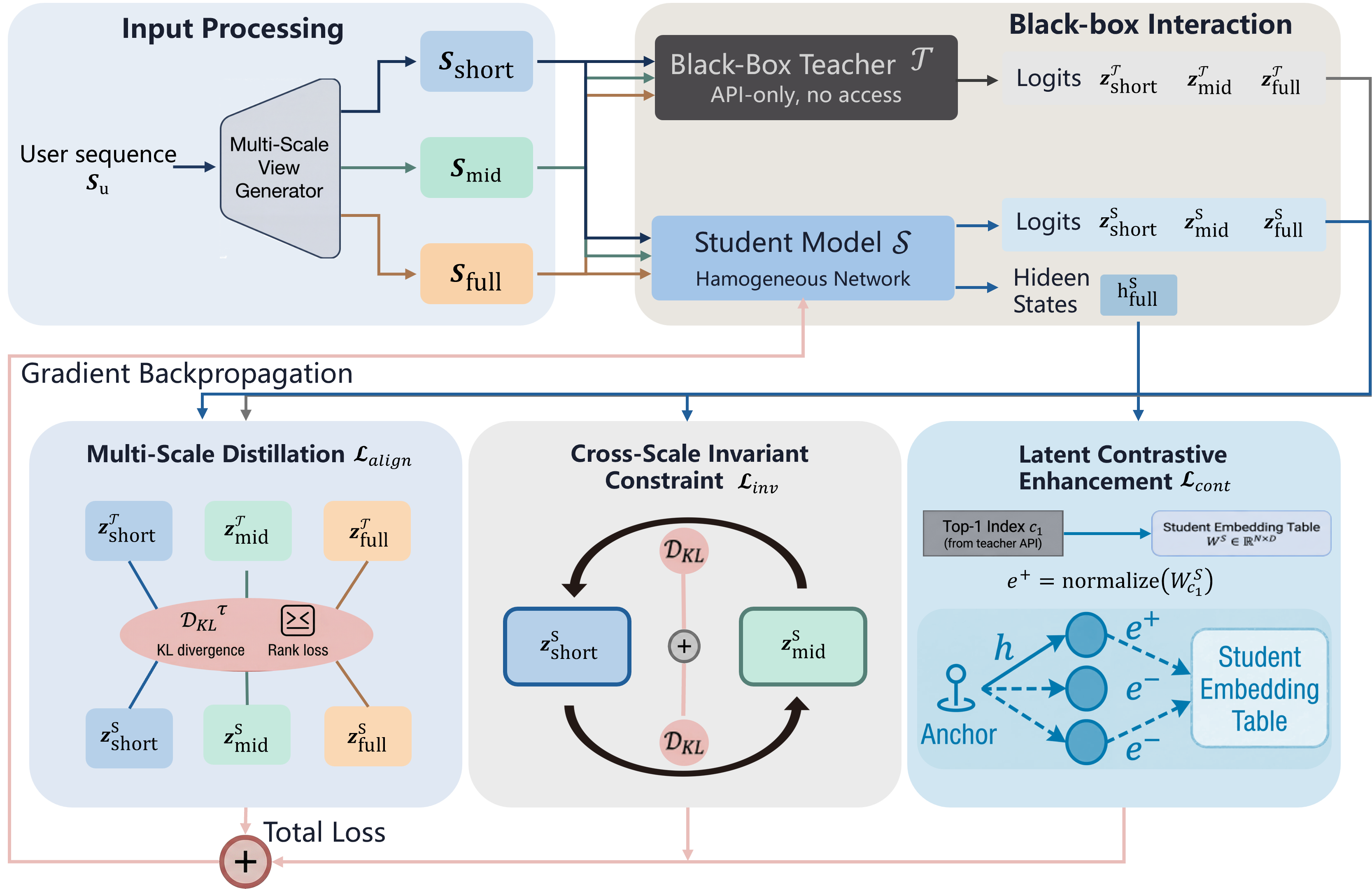}
    \caption{The  framework of BAHSD. A multi-scale view generator probes the teacher API without internal access, while three loss modules regulate gradient flow: $\mathcal{L}_{align}$ transfers multi-scale knowledge, $\mathcal{L}_{inv}$ filters noise via temporal consistency, and $\mathcal{L}_{cont}$ enhances latent discriminability. }
    \label{fig:framework}
\end{figure}

BAHSD is a unified distillation framework that operates under strict black-box constraints without explicit user stratification. It integrates three complementary components to jointly address preference solidification in head users and discriminative vacuum in tail users. Through multi-scale sequence probing, the framework implicitly perceives signal reliability and adaptively adjusts its optimization strategy, enhancing discriminative capability for degraded signals while preserving fine-grained knowledge transfer for high-quality signals. This design transforms black-box distillation into an adaptive, pathology-aware knowledge refinement process, as illustrated in Fig.~\ref{fig:framework}.

\subsection{Notations}
We define the set of users as $\mathcal{U}=\{u_1, \dots, u_M\}$ and items as $\mathcal{I}=\{i_1, \dots, i_N\}$. For a user $u \in \mathcal{U}$, the chronologically ordered interaction sequence is denoted as $S_u = [i_1, \dots, i_L] \in \mathcal{I}^L$, where $L = |S_u|$ (sequences with $L<5$ are filtered). 
\begin{itemize}
    \item \textbf{Teacher Model ($\mathcal{T}$)}: A black-box API outputting logits $\mathbf{z}^\mathcal{T}(S) \in \mathbb{R}^N$. The predictive distribution is $\mathbf{p}^\mathcal{T}(S_u) = \sigma(\mathbf{z}^\mathcal{T}(S_u))$, where $\sigma(\cdot)$ is the softmax function.
    \item \textbf{Student Model ($\mathcal{S}$)}: A lightweight homogeneous architecture. It outputs logits $\mathbf{z}^\mathcal{S}(S_u)$ and hidden states $\mathbf{H}^\mathcal{S}(S_u) \in \mathbb{R}^{L \times D}$.
    \item \textbf{Probing Views}: $S_p$ denotes the multi-scale view $p \in \{short, mid, full\}$, e.g., corresponding to truncated prefixes (50\%, 75\%, 100\%) of $S_u$.
    \item \textbf{Latent Representation}: $\mathbf{h}_p^\mathcal{S} = \text{L2\_Norm}(\mathbf{H}^\mathcal{S}(S_p)) \in \mathbb{R}^D$ represents the aggregated global representation of view $S_p$.
\end{itemize}

\subsection{Multi-scale Semantic Alignment ($\mathcal{L}_{align}$)}
To extract knowledge at different granularities, we define a joint loss for each view $p$:
\begin{equation}
\mathcal{L}_{align} = \sum_{p \in \{short, mid, full\}} \lambda_p \left[ \alpha \cdot \mathcal{L}_{rank}(\mathbf{z}^\mathcal{T}_p, \mathbf{z}^\mathcal{S}_p) + (1-\alpha) \cdot \mathcal{L}_{KL}^{(p)} \right] 
\end{equation}
where $\mathcal{L}_{rank}$ denotes the pairwise Bayesian Personalized Ranking (BPR) loss and $\mathcal{L}_{KL}^{(p)}$ represents the temperature-scaled Kullback-Leibler divergence:
\begin{equation}
 \mathcal{L}_{KL}^{(p)} = \tau_{kd}^2 \cdot \text{KL}\!\left( \sigma(\mathbf{z}^\mathcal{T}_p / \tau_{kd}) \parallel \sigma(\mathbf{z}^\mathcal{S}_p / \tau_{kd}) \right) 
\end{equation}
A large global temperature $\tau_{kd}$ enables adaptive gradient adjustment that responds to signal heterogeneity. For head users exhibiting preference solidification, $\tau_{kd}$ smooths extreme probability peaks to expose dark knowledge, allowing $\mathcal{L}_{KL}$ to dominate fine-grained knowledge transfer. Conversely, for tail users where distributions collapse, the temperature scaling drives both teacher and student distributions toward uniformity, which naturally attenuates the KL divergence gradient. In this regime, the model adaptively transitions to $\mathcal{L}_{rank}$-dominated optimization, focusing on robust ranking learning rather than matching uninformative distributions.

\subsection{Cross-scale Invariance Constraint ($\mathcal{L}_{inv}$)}
Empirical analysis reveals that tail users often exhibit deceptive false confidence due to sparse and homogeneous interactions. To filter such transient perturbations, we introduce a self-supervised symmetric KL divergence that constrains the student model's output consistency across short-scale views:
\begin{equation}
    \mathcal{L}_{inv} = \frac{\tau_{\text{kd}}^2}{2} \left[ \text{KL}(\mathbf{p}_{\text{short}}^\mathcal{S} \| \mathbf{p}_{\text{mid}}^\mathcal{S}) + \text{KL}(\mathbf{p}_{\text{mid}}^\mathcal{S} \| \mathbf{p}_{\text{short}}^\mathcal{S}) \right]
\end{equation}
where $\mathbf{p}^\mathcal{S} = \text{softmax}(\mathbf{z}^\mathcal{S}(S_p) / \tau_{\text{kd}})$. Genuine user intent evolves smoothly over time, whereas pseudo-sharpness, a form of false confidence arising from sparse data, exhibits significant distributional shifts when the input sequence is truncated. When $\mathcal{L}_{inv}$ captures such divergence, it generates strong penalty gradients that guide the student model back toward a more robust representation space. Notably, this constraint excludes the full sequence $S_{\text{full}}$ to preserve its capacity for capturing recent concept drift while ensuring stable and reliable representations for historical intents. This design complements the other loss components by explicitly filtering noise at the distribution level, thereby reinforcing the framework's overall robustness to signal heterogeneity.

\subsection{Sequence-Item Latent Contrastive Enhancement ($\mathcal{L}_{cont}$)}
To mitigate preference solidification in head users and enhance semantic discriminability for tail users, we introduce a latent-space contrastive module based on InfoNCE. Transformer-based models such as SASRec and BERT4Rec share a structural property where the sequence representation $\mathbf{h}_{\text{full}}^{\mathcal{S}}$ and the item embedding matrix $\mathbf{W}^{\mathcal{S}}$ lie in the same semantic manifold, enabling direct sequence-item alignment:

\begin{equation}
    \mathcal{L}_{cont} = - \log \frac{\exp(\text{sim}(\mathbf{h}_{\text{full}}^\mathcal{S}, \mathbf{e}^+) / \tau_{\text{cl}})}{\exp(\text{sim}(\mathbf{h}_{\text{full}}^\mathcal{S}, \mathbf{e}^+) / \tau_{\text{cl}}) + \sum_{k=1}^{N_{\text{neg}}} \exp(\text{sim}(\mathbf{h}_{\text{full}}^\mathcal{S}, \mathbf{e}_k^-) / \tau_{\text{cl}})}
\end{equation}

All components adhere to black-box constraints. The anchor $\mathbf{h}_{\text{full}}^{\mathcal{S}}$ is the L2-normalized final hidden state of the student for the full sequence. The positive sample $\mathbf{e}^+$ is obtained by extracting the top-1 item index $c_1$ from the teacher's logits and mapping it to the student's embedding matrix via $\mathbf{e}^+ = \text{normalize}(\mathbf{W}^{\mathcal{S}}[c_1])$. Negative samples are uniformly drawn from the student's embedding space, with $N_{neg} = 256$.

The temperature parameter $\tau_{\text{cl}}$ controls geometric repulsion in the latent space with two complementary benefits. It mitigates head-user representation collapse by enlarging inter-item distances to alleviate preference solidification, and it addresses tail-user discriminative vacuum through negative sampling that establishes well-defined item boundaries.

\subsection{Joint Optimization and Gradient Dynamics}
The overall optimization objective for the student model is formulated as a weighted combination of the three proposed components:
\begin{equation}
    \mathcal{L}_{total} = \lambda_A \mathcal{L}_{align} + \lambda_B \mathcal{L}_{inv} + \lambda_C \mathcal{L}_{cont}
\end{equation} 
where the hyper-parameters $\lambda_A$, $\lambda_B$, and $\lambda_C$ balance the contribution of each loss term, with dataset- and architecture-specific values provided in the experimental section.


\section{Experiments}
\label{sec:experiments}
To evaluate the effectiveness of BAHSD and its adaptive hierarchical distillation loss, we conduct comprehensive experiments aimed at answering the following research questions (RQs):
\begin{itemize}
\item \textbf{RQ1:} How does BAHSD compare with state-of-the-art black-box knowledge distillation baselines in terms of overall recommendation accuracy?
\item \textbf{RQ2:} How effectively does BAHSD adapt to signal heterogeneity on both sparse   and dense   datasets?
\item \textbf{RQ3:} What are the individual contributions of each loss component to the overall performance?
\item \textbf{RQ4:} How robust is BAHSD to key hyperparameters, such as truncation ratios and temperature coefficients?
\end{itemize}

\subsection{Experimental Setup}
\subsubsection{Datasets}
We conduct experiments on two public benchmarks: Amazon Beauty\cite{ni2019justifying}, a sparse dataset containing approximately 620K interactions, and MovieLens-1M\cite{harper2015movielens}, a dense dataset with approximately 1M interactions. Both datasets are preprocessed by retaining only users with at least five interactions, following standard practice, and we adopt a leave-one-out evaluation protocol for model assessment.
\begin{table}[tp]
  \centering
  \caption{Parameter Settings in the Distillation Phase}
  \label{tab:distillation_params}
  \begin{tabular}  {lccccccccccc}
    \toprule
    Dataset & Backbone & $S_{short}$ & $S_{mid}$ & $\lambda_{short}$ & $\lambda_{mid}$ & $\lambda_{full}$ & $\lambda_{A}$ & $\lambda_{B}$ & $\lambda_{C}$ & $\tau_{kd}$ & $\tau_{cl}$ \\
    \midrule
    \multirow{2}{*}{Amazon Beauty}  & BERT4Rec  & 0.4 & 0.8 & 1.0 & 1.2 & 0.8 & 1  & 1.2 & 0.3       & 1.0  & 0.1                    \\
   & SASRec    & 0.25             & 0.5               & 1.2                 & 1.2                  & 1.8     & 1            & 0.8                  & 0.1                  & 4.0                & 0.1                    \\
   \midrule
    \multirow{2}{*}{MovieLens-1M}  & BERT4Rec    & 0.5              & 0.75              & 1.0                 & 1.2                  & 1.8       & 1          & 0.3                  & 1.0                  & 2.0                & 0.1                    \\
     & SASRec      & 0.5              & 0.75              & 1.0                 & 1.2                  & 1.8       & 1          & 0.8                  & 0.3                  & 2.0                & 0.1                    \\
    \bottomrule
  \end{tabular}
\end{table}

\begin{table}[tp]
  \centering
  \caption{Overall distillation performance in different metricsl, with bold and
underlined entries denoting optimal and suboptimal performance, respectively. ABKD-1: $\alpha=1.2, \beta=1.2$; ABKD-2: $\alpha=0.5, \beta=1.5$; ABKD-3: $\alpha=1.5, \beta=0.5$.}
  \label{tab:overall_perf}
  \scriptsize  
  \renewcommand{\arraystretch}{1.05}  
  \setlength{\tabcolsep}{2pt}         
  \begin{tabular}{lcccccccc}
    \toprule
    \multirow{2}{*}{Method} & \multicolumn{4}{c}{Beauty} & \multicolumn{4}{c}{MovieLens-1M} \\
    \cmidrule(lr){2-5} \cmidrule(lr){6-9}
    & \multicolumn{2}{c}{BERT4Rec} & \multicolumn{2}{c}{SASRec} & \multicolumn{2}{c}{BERT4Rec} & \multicolumn{2}{c}{SASRec} \\
    \cmidrule(lr){2-3} \cmidrule(lr){4-5} \cmidrule(lr){6-7} \cmidrule(lr){8-9}
    & Recall@10 & NDCG@10 & Recall@10 & NDCG@10 & Recall@10 & NDCG@10 & Recall@10 & NDCG@10 \\  
    \midrule
    Teacher & 0.493 & 0.321 & 0.493 & 0.341 & 0.794 & 0.590 & 0.827 & 0.605 \\
    \cmidrule(lr){1-9}     
    DFME    & 0.265 & 0.188 & 0.483 & 0.333 & \underline{\text{0.775}} & 0.559 & 0.802 & 0.602 \\
    ME-MIA  & 0.267 & 0.190 & 0.486 & 0.334 & 0.773 & \underline{\text{0.560}} & \textbf{0.820} & \underline{\text{0.594}} \\
    UnKD    & 0.249 & 0.147 & 0.206 & 0.105 & 0.609 & 0.557 & 0.616 & 0.559 \\
    DHKD    & 0.210 & 0.109 & 0.137 & 0.069 & 0.600 & 0.551 & 0.598 & 0.548 \\
    ABKD-1  & \underline{\text{0.355}} & \underline{\text{0.212}} & 0.432 & 0.263 & 0.742 & 0.508 & 0.792 & 0.559 \\
    ABKD-2  & 0.271 & 0.188 & 0.441 & 0.371 & 0.715 
    & 0.493 & 0.786 & 0.561 \\
    ABKD-3  & 0.266 & 0.181 & 0.435 & 0.263 & 0.723 & 0.495 & 0.791 & 0.556 \\
    CDBCF   & 0.257 & 0.186 & \underline{\text{0.488}} & \underline{\text{0.337}} & \textbf{0.782} & \textbf{0.571} & \underline{\text{0.818}} & \textbf{0.595} \\
    \textbf{BAHSD}    & \textbf{0.465} & \textbf{0.297} & \textbf{0.525} & \textbf{0.358} & 0.767 & 0.541 & 0.816 & 0.580 \\
    \bottomrule
  \end{tabular}
\end{table}



\subsubsection{Backbone Models}
We employ two representative architectures as backbone models: SASRec \cite{kang2018self}, which utilizes unidirectional self-attention, and BERT4Rec \cite{sun2019bert4rec}, which employs bidirectional self-attention. These models represent the mainstream paradigms for modeling temporal user behavior. All distillation experiments are conducted in a homogeneous setting, where both the teacher and student models share identical architectures.

\begin{figure}[!htb]
  \centering
  \subfloat[Amazon beauty on BERT4Rec]{
    \includegraphics[width=0.45\textwidth]{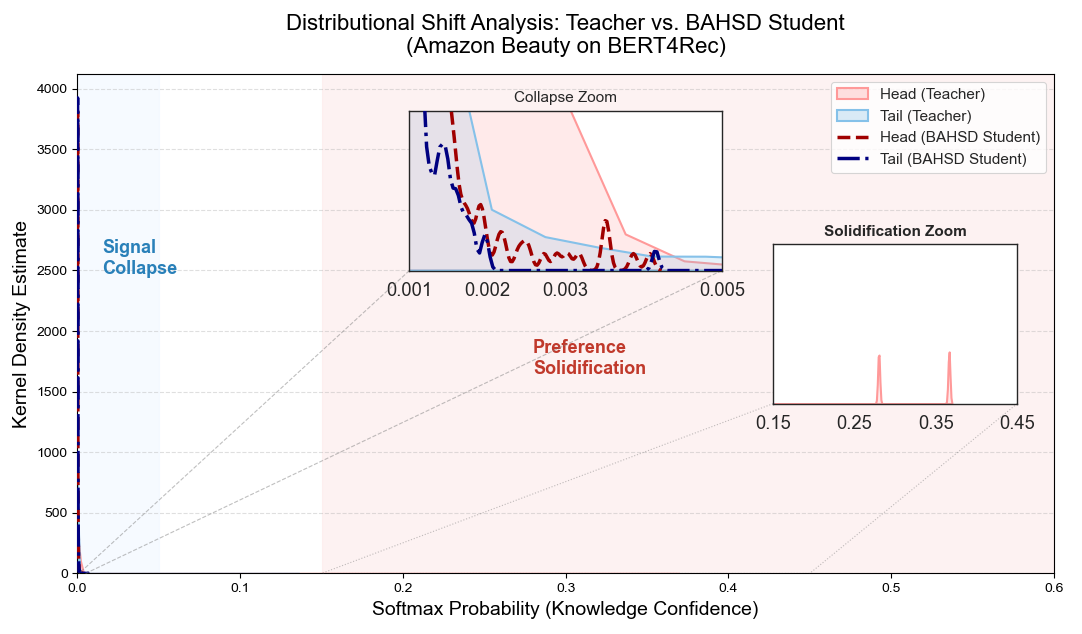}
    \label{subfig:beauty-bert}
  }
  \hfill  
  \subfloat[Amazon beauty on SASRec]{
    \includegraphics[width=0.45\textwidth]{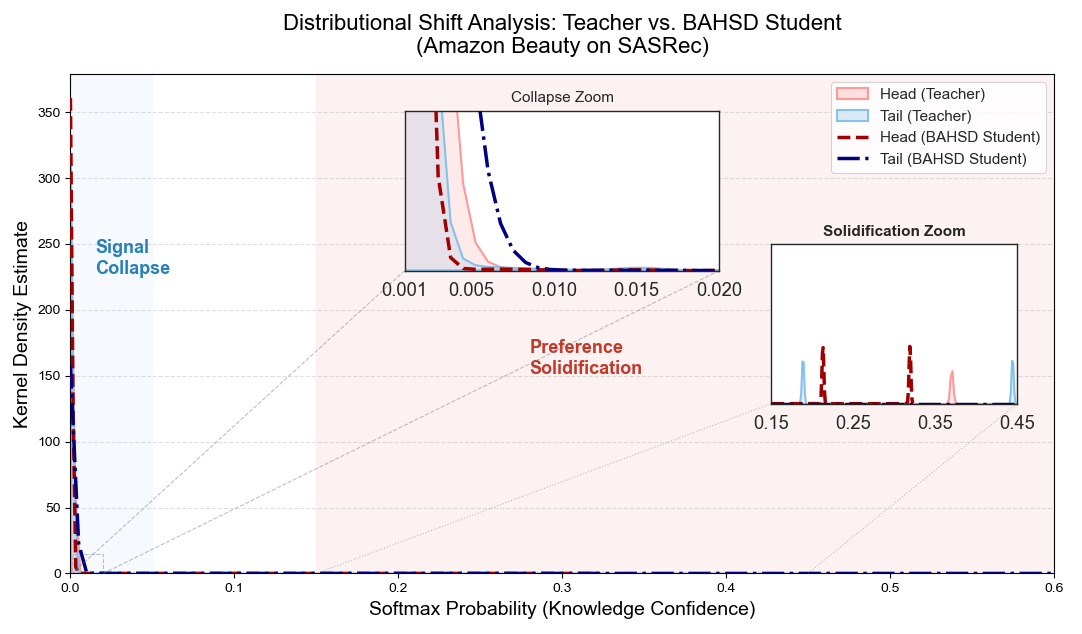}
    \label{subfig:beauty-sas}
  }\\[0pt]  
  \subfloat[MovieLens-1M on BERT4Rec]{
    \includegraphics[width=0.45\textwidth]{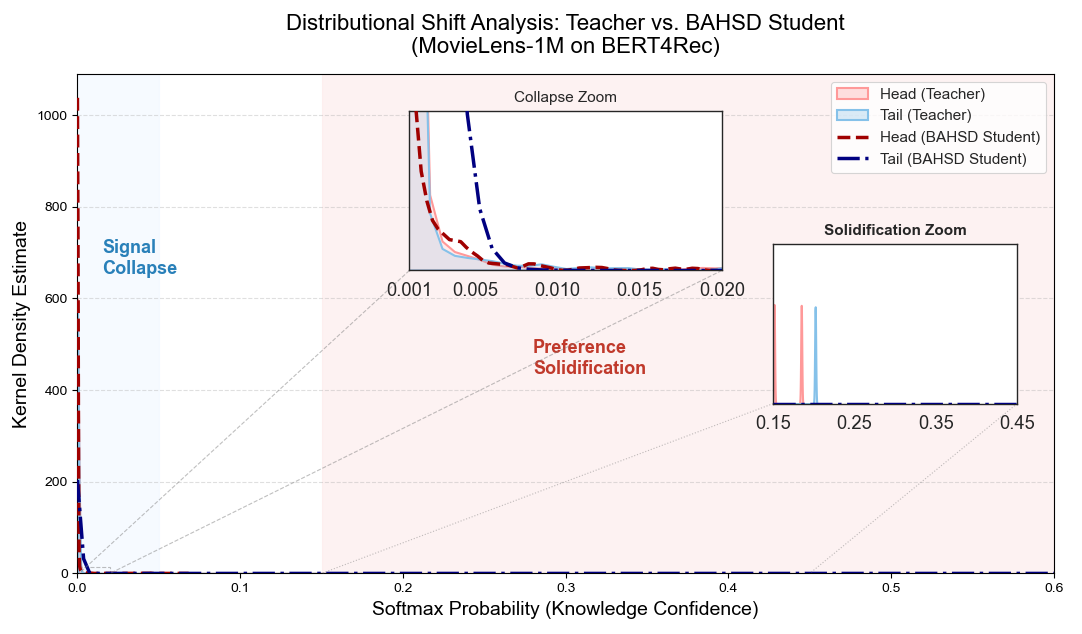}
    \label{subfig:ml1m-bert}
  }
  \hfill
  \subfloat[MovieLens-1M on SASRec]{
    \includegraphics[width=0.45\textwidth]{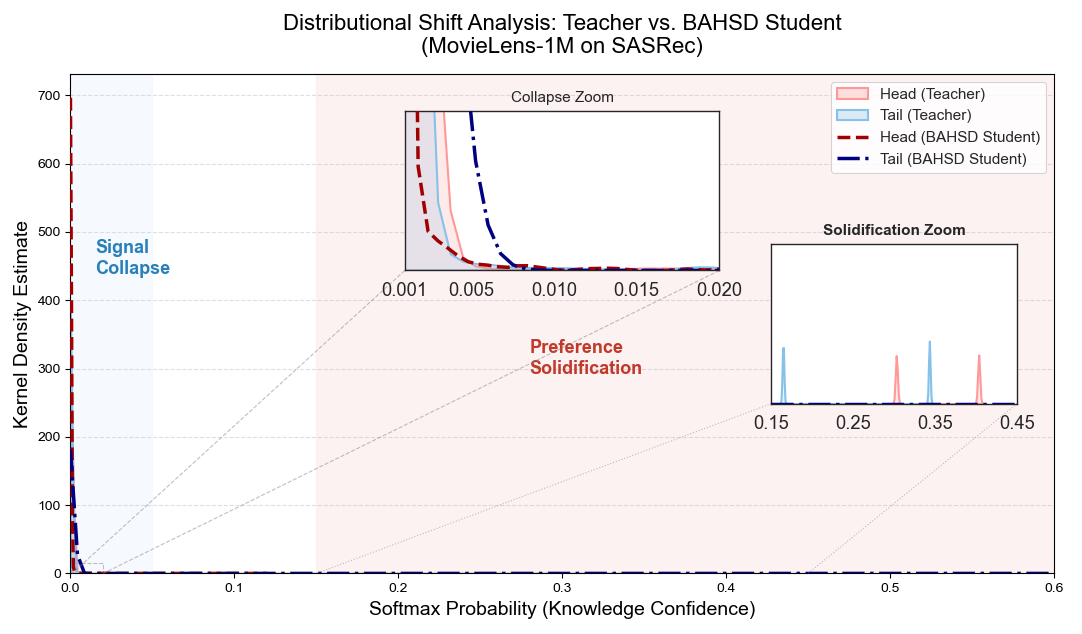}
    \label{subfig:ml1m-sas}
  }
  \caption{Distribution shift analysis between teacher and student models.}
  \label{fig:dist_shift}
\end{figure}

\subsubsection{Baselines}
We compare BAHSD against the following strong baselines:
\begin{itemize}
    \item \textbf{DFME} \cite{yue2021black}: Extracts black-box models through synthetic sequence generation and pairwise ranking distillation, aligning the order of top-k items via marginal ranking loss.
    \item \textbf{ME-MIA} \cite{zhu2023membership}: Optimizes extraction using a combination of BPR ranking loss and Hinge loss, achieving enhanced performance through engineering refinements.
    \item \textbf{UnKD} \cite{chen2023unbiased}: Mitigates popularity bias by stratifying items into groups and optimizing ranking consistency within each group.
    \item \textbf{DHKD} \cite{yang2025dual}: Employs a dual-head distillation strategy to alleviate gradient conflicts between cross-entropy and logit-level losses.
    \item \textbf{ABKD} \cite{wang2025abkd}: Utilizes $\alpha$-$\beta$ divergence as a smooth interpolation between forward and reverse KL divergences to balance mode-covering and mode-seeking effects.
    \item \textbf{CDBCF} \cite{zhang2026cognitive}: Implements distribution reconstruction via attention-decay theory for black-box model extraction.
\end{itemize}
\subsubsection{Implementation Details}
The SASRec and BERT4Rec backbone models are configured with 100-dimensional embeddings, two Transformer encoder layers, and two self-attention heads, each with a dimensionality of 100, ensuring architectural consistency between the teacher and student models. For distillation, we employ the Adam optimizer with a learning rate of 0.001, a batch size of 512, and a total of 100 training epochs. During testing, we adopt a sampling strategy that pairs one positive item with 99 negative items per test instance, and we evaluate distillation performance using NDCG@10 and Recall@10. All experiments are conducted on an NVIDIA L40 GPU with PyTorch 2.9.1 and CUDA 12.4. The specific loss weights for each dataset and backbone combination are provided in Table~\ref{tab:distillation_params}.

\begin{table}[tp]
\centering
\small
\caption{Tier-wise performance comparison on \textbf{Amazon Beauty}, with bold and
underlined entries denoting optimal and suboptimal performance, respectively.}
\label{tab:beauty_performance}
\begin{tabular}{llcccccccc}
\toprule
\multirow{2}{*}{Backbone} & \multirow{2}{*}{Method} & \multicolumn{4}{c}{Recall@10} & \multicolumn{4}{c}{NDCG@10} \\
\cmidrule(lr){3-6} \cmidrule(lr){7-10}
& & Overall & Head & Mid & Tail & Overall & Head & Mid & Tail \\
\midrule
\multirow{5}{*}{BERT4Rec} & Teacher & 0.493 & 0.527 & 0.500 & 0.478 & 0.321 & 0.357 & 0.322 & 0.308 \\
\cmidrule(lr){2-10} 
& DFME & 0.265 & \underline{\text{0.289}} & \underline{\text{0.273}} & 0.255 & 0.188 & \underline{\text{0.205}} & \underline{\text{0.193}} & 0.181 \\
& ME-MIA & \underline{\text{0.267}} & 0.281 & 0.271 & \underline{\text{0.260}} & \underline{\text{0.190
}}& 0.204 & 0.191 & \underline{\text{0.185}} \\
& CDBCF & 0.257 & 0.279 & 0.267 & 0.247 & 0.186 & 0.203 & 0.191 & 0.178 \\
& \textbf{BAHSD} & \textbf{0.465} & \textbf{0.514} & \textbf{0.455} & \textbf{0.448} & \textbf{0.297} & \textbf{0.346} & \textbf{0.287} & \textbf{0.280} \\
\midrule
\multirow{5}{*}{SASRec} & Teacher & 0.493 & 0.547 & 0.493 & 0.474 & 0.341 & 0.395 & 0.340 & 0.322 \\
\cmidrule(lr){2-10}
& DFME & 0.483 & 0.533 & 0.480 & 0.466 & 0.333 & 0.386 & 0.328 & 0.315 \\
& ME-MIA & 0.486 & 0.532 & 0.484 & 0.466 & 0.334 & 0.383 & 0.332 & 0.315 \\
& CDBCF & \underline{\text{0.488}} & \underline{\text{0.536}} & \underline{\text{0.493}} & \underline{\text{0.469}} & \underline{\text{0.337}} & \underline{\text{0.388}} & \textbf{0.377} & \underline{\text{0.318}} \\
& \textbf{BAHSD} & \textbf{0.525} & \textbf{0.573} & \textbf{0.516} & \textbf{0.505} & \textbf{0.358} & \textbf{0.417} & \underline{\text{0.345}} & \textbf{0.338} \\
\bottomrule
\end{tabular}
\end{table}

\subsection{Experimental Results and Analysis}

\begin{table*}[tp]
\centering
\caption{Tier-wise performance comparison on \textbf{MovieLens-1M}, with bold and
underlined entries denoting optimal and suboptimal performance, respectively.}
\label{tab:ml1m_performance}
\begin{tabular}{llcccccccc}
\toprule
\multirow{2}{*}{Backbone} & \multirow{2}{*}{Method} & \multicolumn{4}{c}{Recall@10} & \multicolumn{4}{c}{NDCG@10} \\
\cmidrule(lr){3-6} \cmidrule(lr){7-10}
& & Overall & Head & Mid & Tail & Overall & Head & Mid & Tail \\
\midrule
\multirow{5}{*}{BERT4Rec} & Teacher & 0.794 & 0.689 & 0.745 & 0.844 & 0.590 & 0.478 & 0.549 & 0.641 \\
\cmidrule(lr){2-10}

& DFME & \underline{\text{0.775}} & \underline{\text{0.650}} & 0.728 & \underline{\text{0.837}} & 0.559 & \underline{\text{0.439}} & 0.518 & 0.620 \\
& ME-MIA & 0.773 & \textbf{0.652} & 0.722 & 0.830 & \underline{\text{0.560}} & \textbf{0.440} & 0.515 & 0.614 \\
& CDBCF & \textbf{0.782} & 0.644 & \underline{\text{0.731}} & \underline{\text{0.837}} & \textbf{0.571} & 0.433 & \underline{\text{0.523}} & \underline{\text{0.622}} \\
& \textbf{BAHSD} & 0.767 & 0.610 & \textbf{0.783} & \textbf{0.866} & 0.541 & 0.380 & \textbf{0.557} & \textbf{0.650} \\
\midrule
\multirow{5}{*}{SASRec} & Teacher & 0.827 & 0.713 & 0.791 & 0.875 & 0.605 & 0.487 & 0.562 & 0.658 \\
\cmidrule(lr){2-10}

& DFME & 0.802 & 0.661 & 0.749 & 0.855 & \textbf{0.602} & 0.443 & 0.535 & 0.641 \\
& ME-MIA & \textbf{0.820} & 0.694 & \underline{\text{0.778}} & \underline{\text{0.872}} & 0.594 & \underline{\text{0.461}} & \underline{\text{0.556}} & \underline{\text{0.651}} \\
& CDBCF & \underline{\text{0.818}} & \underline{\text{0.699}} & \underline{\text{0.778}} & 0.869 & \underline{\text{0.595}} & \textbf{0.472} & 0.553 & 0.650 \\
& \textbf{BAHSD} & 0.816 & \textbf{0.701} & \textbf{0.825} & \textbf{0.879} & 0.580 & 0.455 & \textbf{0.589} & \textbf{0.673} \\
\bottomrule
\end{tabular}
\end{table*}

\subsubsection{Overall Distillation Performance (RQ1)}
Table~\ref{tab:overall_perf} reports the distillation performance across different scenarios. BAHSD consistently outperforms all baselines. Notably, on Amazon Beauty with SASRec, BAHSD achieves a relative improvement of 81.38\% for tail users and even surpasses the teacher model by 4.75\%.
Figure~\ref{fig:dist_shift} illustrates the distribution evolution during distillation, revealing two core mechanisms of BAHSD. First, it mitigates preference solidification in head users by softening the teacher's overconfident peaks ($p>0.15$), transforming rigid pattern memorization into generalized preference modeling. Second, it recovers discriminative power for tail users by reconstructing clear semantic boundaries in the discriminative vacuum region ($p<0.05$), compensating for the structural information loss inherent in standard KL divergence-based distillation. We also observe that BAHSD achieves better performance with SASRec than BERT4Rec, suggesting that bidirectional masking poses greater challenges for black-box distillation.
\subsubsection{ User-Tier Performance Analysis (RQ2)}
Tier-wise comparisons in Tables~\ref{tab:beauty_performance} and~\ref{tab:ml1m_performance} demonstrate that BAHSD effectively adapts to signal heterogeneity across user strata. On the sparse Amazon Beauty dataset, BAHSD achieves substantial improvements for tail users, with relative Recall@10 gains reaching 81.38\%, attributed to InfoNCE-based structural boundary reconstruction that suppresses stochastic noise and compensates for the discriminative vacuum inherent in sparse interactions. On the dense MovieLens-1M dataset, BAHSD maintains stable head-user performance while delivering significant gains for mid and tail tiers, including a 19.4\% relative Recall@10 improvement for tail users. The framework intentionally decouples head users from the teacher's over-concentrated distributions to avoid inheriting localized overfitting patterns, fostering generalized preference boundaries, while cross-scale consistency constraints refine the representation space for mid and tail users to ensure structural alignment despite incomplete teacher context.
\begin{table*}[!htb]
\centering
\caption{Ablation study of BAHSD components across datasets and backbones. Bold and underlined entries indicate optimal and suboptimal performance.}
\label{tab:ablation}
\resizebox{\textwidth}{!}{
\begin{tabular}{lcccccccc}
\toprule
\multirow{2}{*}{\textbf{Configuration}} & \multicolumn{4}{c}{\textbf{Amazon Beauty (Sparse)}} & \multicolumn{4}{c}{\textbf{MovieLens-1M (Dense)}} \\
\cmidrule(lr){2-5} \cmidrule(lr){6-9}
 & \multicolumn{2}{c}{BERT4Rec} & \multicolumn{2}{c}{SASRec} & \multicolumn{2}{c}{BERT4Rec} & \multicolumn{2}{c}{SASRec} \\
\cmidrule(lr){2-3} \cmidrule(lr){4-5} \cmidrule(lr){6-7} \cmidrule(lr){8-9}
 & Recall@10 & NDCG@10 & Recall@10 & NDCG@10 & Recall@10 & NDCG@10 & Recall@10 & NDCG@10 \\
\midrule
\textbf{Full Model} & \underline{\text{0.465}} & \underline{\text{0.297}} & \textbf{0.525} & \textbf{0.358} & \underline{\text{0.767}} & \textbf{0.541} & \textbf{0.816} & 0.580 \\
\cmidrule(lr){1-9}
w/o $\mathcal{L}_{short}$  & 0.457  & 0.289  & 0.521  & 0.354  & 0.753  & 0.523  & 0.812  & 0.579  \\
w/o $\mathcal{L}_{mid}$  & 0.456  & 0.288  & \textbf{0.525}  & \underline{\text{0.356}}  & 0.765  & 0.533  & 0.811  & 0.576  \\
w/o $\mathcal{L}_{full}$  & \textbf{0.481}  & \textbf{0.307}  & 0.521  & 0.353  & 0.460  & 0.255  & 0.778  & 0.544  \\
w/o $\mathcal{L}_{inv}$  & 0.452  & 0.288  & 0.523  & 0.355  & \textbf{0.771}  & \underline{\text{0.540}}  & \underline{\text{0.815}}  & \textbf{0.586}  \\
w/o $\mathcal{L}_{cont}$  & 0.422  & 0.269  & \underline{\text{0.524}}  & 0.355  & 0.757  & 0.535  & 0.810  & \underline{\text{0.582}}  \\
\bottomrule
\end{tabular}
}
\end{table*}

\begin{figure}[!htb]
  \centering
  \includegraphics[width=\textwidth]{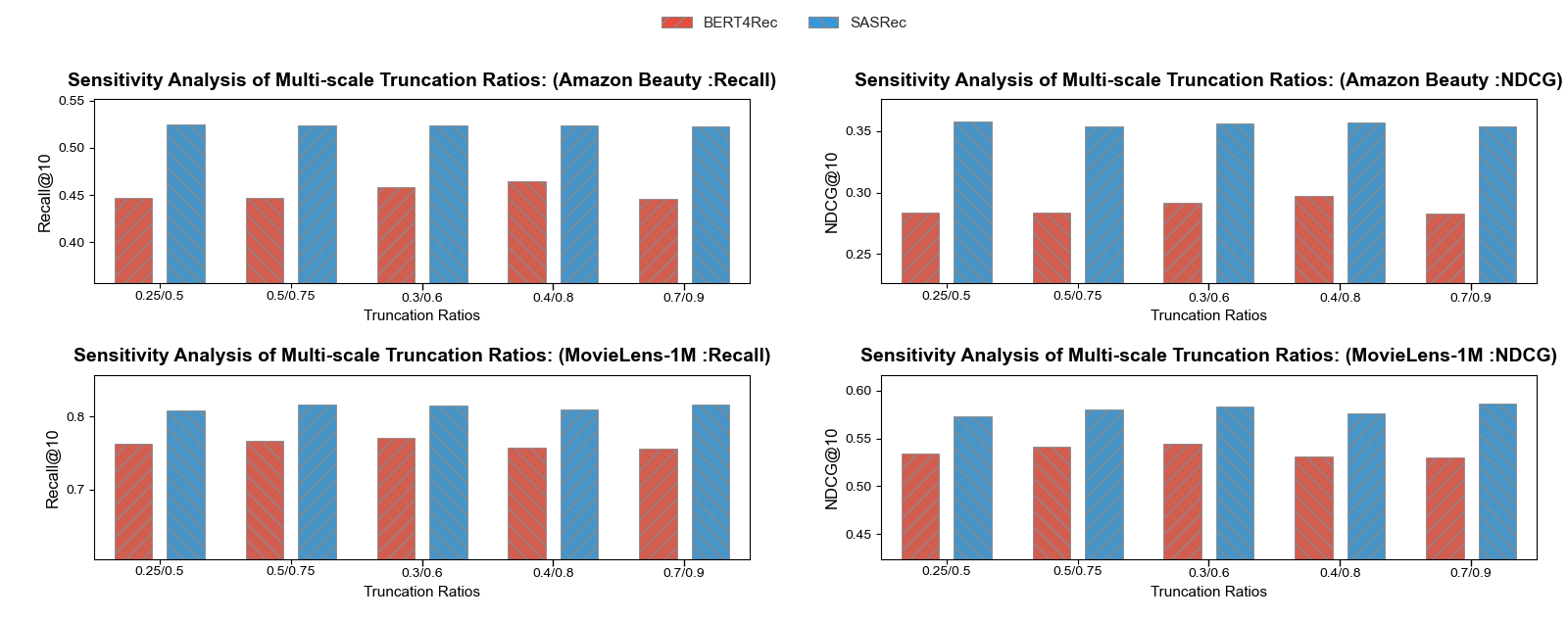}
  \caption{Impact of truncation ratios   on distillation performance.}
  \label{fig:truncation_ratio}
\end{figure}
\begin{figure}[tp]
  \centering
  \includegraphics[width=\textwidth]{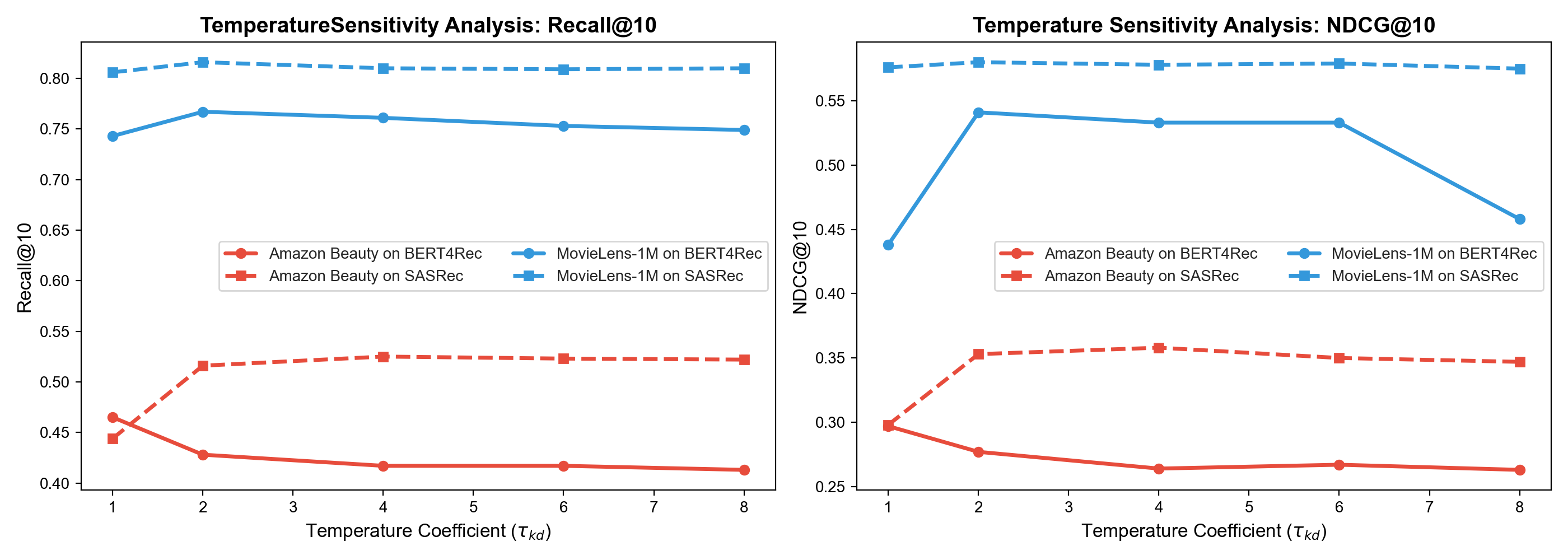}
  \caption{Impact of distillation temperature $\tau_{kd}$ on distillation performance.}
  \label{fig:temperature}
\end{figure}

\subsubsection{Ablation Studies (RQ3)}
Ablation results in Table~\ref{tab:ablation} demonstrate the synergistic contribution of each component, as removing any single module degrades performance. The contrastive enhancement module $\mathcal{L}_{cont}$ is essential for constructing discriminative boundaries and suppressing tail noise, with its removal causing a 9.3\% Recall@10 drop on Amazon Beauty with BERT4Rec, though its impact lessens in dense scenarios. The cross-scale invariance constraint $\mathcal{L}_{inv}$ acts as an effective denoiser for sparse datasets by filtering transient perturbations from tail users, yet its effect becomes marginal in dense settings where it may over-regularize intent dynamics. The multi-scale alignment loss $\mathcal{L}_{align}$ drives high-fidelity knowledge transfer in dense scenarios by preserving temporal dependencies, while in sparse settings it provides basic alignment and cooperates with $\mathcal{L}_{cont}$. These results confirm that a one-size-fits-all distillation strategy is suboptimal and that adaptive component synergy is essential for handling signal heterogeneity.

\subsubsection{Impact of Multi-Scale Truncation Ratios (RQ4)} Sensitivity experiments on five truncation ratio pairs (Fig.~\ref{fig:truncation_ratio}) show that optimal ratios depend primarily on dataset sparsity rather than architecture. For sparse Amazon Beauty, the optimal ratios are 0.4 for short sequences and 0.8 for mid sequences, where increasing the mid ratio from 0.5 to 0.8 yields a 4.58\% NDCG@10 gain for BERT4Rec. For dense MovieLens-1M, optimal ratios shift to 0.3 and 0.6, where reducing the mid ratio from 0.9 to 0.6 improves NDCG@10 by 2.83\%. Extreme ratios, such as 0.25/0.5 or 0.7/0.9, consistently degrade performance. The robust optimal range is 0.3-0.4 for short sequences and 0.6-0.8 for mid sequences, with sparser datasets favoring the higher end.
\subsubsection{Impact of Distillation Temperature (RQ4)} Experiments with $\tau_{kd}$ ranging from 1 to 8 (Fig.~\ref{fig:temperature}) reveal that optimal temperatures are architecture-dependent. BERT4Rec favors lower settings, such as $\tau_{kd}=1$ on Beauty and $\tau_{kd}=2$ on MovieLens-1M, to focus on core preferences, while SASRec achieves optimal results at moderate values, including $\tau_{kd}=4$ on the sparse dataset and $\tau_{kd}=2$ on the dense dataset, to preserve temporal correlations. Extreme temperatures consistently degrade performance: excessively low $\tau_{kd}$ induces knowledge narrowing that particularly harms sparse datasets, whereas overly high $\tau_{kd}$ leads to knowledge blurring that reduces accuracy on dense datasets. Sparse datasets exhibit narrower optimal temperature ranges with greater sensitivity, while dense datasets demonstrate wider robustness and stability.
\section{Conclusion}
\label{sec:conclusion}
This paper introduces BAHSD, a black-box adaptive distillation framework designed to address signal heterogeneity in sequential recommendation. Through empirical analysis, we identify two pathological signal modes, namely preference solidification for head users and information vacuum for tail users, which undermine the effectiveness of uniform distillation objectives. BAHSD tackles these challenges by employing multi-scale consistency probing that implicitly perceives signal reliability and by integrating an adaptive hierarchical loss that dynamically adjusts optimization strategies according to the perceived signal quality. Extensive experiments on both sparse and dense datasets demonstrate that BAHSD consistently outperforms state-of-the-art baselines, achieving substantial improvements for tail users while preserving head-user performance and even surpassing the teacher model in certain scenarios. BAHSD offers a model-agnostic and plug-and-play solution for high-fidelity black-box model extraction.


This work provides a robust, model-agnostic, and plug-and-play solution for the local deployment of industrial sequential recommendation models. Looking forward, our future research will focus on extending BAHSD to scenarios with even higher item-space cardinality and exploring the potential of utilizing federated learning frameworks to further mitigate privacy concerns in black-box distillation.

\subsubsection{Acknowledgments.}
This research was funded by the National Key Research and
Development Program of China (Grant No. 2024YFC3307400), under the "Social Governance and Smart Society Technology Support" key special project.
%
%
%
%

\bibliographystyle{splncs04}
\bibliography{refs}

\end{document}